# Numerical Study of Flow Patterns and Heat Transfer in Mini Twisted Oval Tubes

Amin Ebrahimi [1], Ehsan Roohi [1, *]

1- High Performance Computing Laboratory, Department of Mechanical Engineering, Faculty of Engineering, Ferdowsi University of Mashhad, Mashhad, P.O. Box 91775-1111, Khorasan-Razavi, Iran.

* - Corresponding Author, Email: e.roohi@um.ac.ir, Phone/fax: +985138763304, Address: High Performance Computing Laboratory, Department of Mechanical Engineering, Faculty of Engineering, Ferdowsi University of Mashhad, Mashhad, P.O. Box 91775-1111, Khorasan-Razavi, Iran.

**Abstract**

Flow patterns and heat transfer inside mini twisted oval tubes (TOTs) heated by constant-temperature walls are numerically investigated. Different configurations of tubes are simulated using water as the working fluid with temperature-dependent thermo-physical properties at Reynolds numbers ranging between 500 and 1100. After validating the numerical method with the published correlations and available experimental results, the performance of TOTs is compared to a smooth circular tube. The overall performance of TOTs is evaluated by investigating the thermal-hydraulic performance and the results are analyzed in terms of the field synergy principle and entropy generation. Enhanced heat transfer performance for TOTs is observed at the expense of a higher pressure drop. Additionally, the secondary flow generated by the tube-wall twist is concluded to play a critical role in the augmentation of convective heat transfer, and consequently, better heat transfer performance. It is also observed that the improvement of synergy between velocity and temperature gradient and lower irreversibility cause heat transfer enhancement for TOTs.

# 1. Introduction

Heat exchangers are extensively used in automotive and aerospace industries, refrigeration systems, power systems and air conditioning. The efficiency of heat exchangers could be improved by enhancing the heat transfer coefficient and reducing the pressure drop [1].

There have been sizeable investigations providing insight into different ways of heat transfer enhancement categorized into active and passive methods [2-4]. The active methods require external power such as surface vibration and acoustic or electrical fields, while the passive methods use distinct surface geometries or fluid additives [3, 5, 6]. Twisted oval tubes (TOTs), shown in Figure 1, are one of the passive techniques to enhance the heat transfer [7]. The twisted oval tube heat exchangers (TOTHE) are aimed to enhance the tube side heat transfer and simultaneously decrease the pressure drop of the shell side, thus promoting the efficiency of heat exchangers. Nowadays, TOTHEs are noticeably favored due to their promising performance.

TOTHEs were developed in the 1980s, and ever since many researchers have focused on studying the heat transfer performance and pressure loss inside the tube and its shell side. Bishara et al. [8, 9] numerically investigated the heat transfer performance of laminar flow with Prandtl number ($Pr$) of 3.0 in an elliptical tube with twist ratio of 6 and aspect ratio ($AR$) of about 1.43 at Reynolds numbers ($Re$) lower than 1200. Reporting the promotion of the heat transfer in twisted elliptical tubes, the flow characteristics of a twisted elliptical tube, which is of great importance, were not considered in their study. Asmantas et al. [10] experimentally studied turbulent air flow inside twisted flat tubes and presented correlations for heat transfer and friction factor for $Re$ of 7000 to 200,000. The mechanism of heat transfer augmentation in the twisted elliptical tube is not yet entirely clear. Yang et al. [7] tested the heat transfer and flow losses characteristics of water flow in twisted elliptical tubes from the laminar to turbulent regime and compared their performance with circular tubes. Observing higher overall heat transfer and pressure drop in a twisted elliptical tube, they reported the low-$Re$ flow as the best operational regime for twisted elliptical tubes. They, in addition, proposed correlations to predict heat transfer coefficient and friction factor for twisted elliptical tubes and experienced the transition to turbulence at lower $Re$ compared to the smooth circular tube. The heat transfer and pressure drop performances of TOTs in turbulent regime have been investigated experimentally and numerically by Tan et al. [11]. Therein, the impact of geometrical factors on the performance of the TOTs are analyzed, where increasing the cross sectional $AR$ or decreasing twist pitch resulted in an increase of both heat transfer coefficient and friction factor. Additionally, they analyzed their results from field synergy point of view.

To explore the overall performance of TOTHEs many studies focused on effects of using twisted oval tubes on shell side. Dzyubenko and coworkers [12-14] studied the turbulent intensity, boundary layer thickness and heat diffusion coefficient experimentally and also presented numerical models to calculate the shell side heat removal and pressure drop performance [15]. Additionally, experimental and numerical studies have been performed to introduce correlations to evaluate heat transfer and pressure loss on the shell side [15, 16]. Tan et al. [17] experimentally studied the heat transfer and pressure losses in a TOTHE on both tube and shell side. They indicated that the efficiency of the TOTHEs is higher at low tube and high shell side flow rate. Tan et al. [18] also implemented a numerical investigation of pressure drop and heat transfer characteristics in the shell side of TOTHE in the turbulent regime.

Better understanding of heat transfer mechanisms in TOTs is of great importance for energy saving and developing more efficient TOTHEs. Although the first law of thermodynamics is always used to study the energy transfer in TOTHEs, the entropy generation analysis could help to understand the mechanism of heat transfer enhancement. Moreover, many researchers have concentrated on entropy generation inside the ducts and tubes for concurrent optimization of heat transfer and fluid flow [19-22].

In the foregoing literatures, most investigations on the heat transfer augmentation in TOTs are based on Nusselt number and friction factor, which are not generally enough to describe the heat transfer enhancement. Moreover, beside the extensive interest in the heat transfer investigations, analyzing the flow pattern and its subsequent effects on the heat transfer have received much less attention. Insomuch as the promising performance of twisted oval tubes and lack of studies on the laminar flow pattern and heat transfer enhancement mechanisms, further investigation of the flow and heat transfer characteristics of twisted oval tubes is indispensable. In the present study, the laminar convective heat transfer in twisted oval tubes with different *AR*s are numerically investigated. These simulations aim to find the effects of the twisted oval tube on the thermo-hydrodynamic behavior of the flow at different Re. Our goal is to evaluate the overall performance of twisted oval tubes and explore, to the best of our knowledge for the first time, the essence of the heat transfer enhancement based on the field synergy principle and entropy generation analysis.

## 2. Model Descriptions

### 2.1. Geometric configurations

Three dimensional simulations are conducted on four different configurations of twisted oval tubes with different cross sectional *AR*s (*a/b*). Figure 1 shows the physical model and relevant geometrical parameters of twisted tubes used in this study. The length of the major and minor axes of the tube cross section is equal to 2a and 2b, respectively. The hydraulic diameter ($D_h$) of all cases considered in this study is five millimeters and the twisted zone length (*S*) is $60D_h$. The computational domain includes three zones: the inlet zone representing the flow developing zone with adiabatic walls, the outlet zone with a length of $10D_h$ and adiabatic walls constructed to avoid any probable back flow streams, and the heated zone containing the heated wall. The length of the inlet zone (*i*) is studied to obtain an almost fully developed flow at the inlet of the twisted zone with minimum length. The twist pitch of the twisted zone is nearly $30D_h$ and the flow direction at the inlet is along the Z direction of Cartesian coordinate.

## 2.2. Mathematical models, governing equations and boundary conditions

Water is used as the working fluids which is modeled to be Newtonian. Laminar and incompressible flow with temperature-dependent thermo-physical properties is considered through the twisted oval tubes with a constant wall temperature. Radiation and gravity effects are neglected in this study. The steady-state conservation equations of mass, momentum and energy can be respectively written as:

$$\nabla \cdot \rho \mathbf{U} = 0 \qquad (1)$$

$$\nabla \cdot (\rho \mathbf{U}\mathbf{U}) = -\nabla p + \nabla \cdot \left[ \mu \left( \nabla \mathbf{U} + (\nabla \mathbf{U})^T \right) \right] \qquad (2)$$

$$\nabla \cdot (\rho c_p T \mathbf{U}) = \nabla \cdot (k \nabla T) \qquad (3)$$

where, $\mathbf{U}$ is the fluid velocity vector and $\rho$, $\mu$, $k$, $T$, $c_p$ and $p$ are density, dynamic viscosity, thermal conductivity, temperature, specific heat capacity and pressure, respectively. The effective temperature-dependent thermo-physical properties of water is summarized and presented in Table 1.

The boundary conditions for the fluid flow at the inlet, outlet and walls are defined by:

Inlet: $u = U_i, \quad v = w = 0, \quad T = T_i = 298.15 \, (K) \qquad (4)$

Outlet: $\dfrac{\partial u}{\partial z} = \dfrac{\partial v}{\partial z} = \dfrac{\partial w}{\partial z} = 0, \quad \dfrac{\partial T}{\partial z} = 0 \qquad (5)$

Adiabatic wall: $u = v = w = 0, \quad k \dfrac{\partial T}{\partial y} = 0 \qquad (6)$

Heated wall: $u = v = w = 0, \quad T = T_{wall} = 348.15 \, (K) \qquad (7)$

where $u$, $v$ and $w$ are velocity components in X, Y and Z direction, respectively. The subscripts of "*i*" and "*wall*" stand for inlet and wall conditions, respectively.

## 2.3. Numerical methods and parameter definitions

The discretization of the computational domain is performed by a second order upwind scheme and a structured non-uniform mesh is generated using the ANSYS Gambit (ANSYS, Inc., Canonsburg, PA, USA), in which the computational grid is refined near the walls to enhance the accuracy and keep the computational cost at bay. All simulations are carried out with the finite volume code ANSYS FLUENT 14.0. The convergence criterion is reached when the scaled residuals of momentum, continuity and energy equations reach to a value lower than $1.0 \times 10^{-10}$, $1.0 \times 10^{-10}$ and $1.0 \times 10^{-12}$, respectively.

For the presentation purposes the following parameters are introduced. The Reynolds number ($Re$) based on the hydraulic diameter ($D_h$) and apparent friction factor ($f$) are defined by the following equations [23].

$$Re = \frac{\rho U_i D_h}{\mu} \quad (8)$$

$$f = \frac{2\Delta p}{\rho U_i^2}\frac{D_h}{L} \quad (9)$$

$$\Delta p = \left(\overline{p_o} - \overline{p_i}\right) \quad (10)$$

$$\overline{p} = \frac{\int p\, dA}{\int dA} \quad (11)$$

where $A$ is the area, $L$ is the heated zone length and the subscript "$o$" belongs to the outlet. Moreover, in chorus with the heat transfer references, the Nusselt number ($Nu$), mean Nusselt number ($Nu_m$) and performance factor ($\eta$) are defined as [24, 25]:

$$Nu = \frac{hD_h}{k} \quad (12)$$

$$h = \frac{Q}{A_{ht}\Delta T} \quad (13)$$

$$Q = \dot{m}c_p\left(\overline{T_o} - \overline{T_i}\right) \quad (14)$$

$$\Delta T = \frac{\left(T_{wall} - \overline{T_i}\right) - \left(T_{wall} - \overline{T_o}\right)}{\ln\left[\left(T_{wall} - \overline{T_i}\right)/\left(T_{wall} - \overline{T_o}\right)\right]} \quad (15)$$

$$\overline{T} = \frac{\int T\rho|\mathbf{U}\cdot d\mathbf{A}|}{\int \rho|\mathbf{U}\cdot d\mathbf{A}|} \quad (16)$$

$$Nu_m = -\frac{D_h}{k_m}\ln\left(\frac{T_{wall} - \overline{T_i}}{T_{wall} - \overline{T_o}}\right)\frac{\dot{m}c_{p,m}}{A_{ht}} \quad (17)$$

$$\eta = \left(Nu_m / Nu_{m,0}\right)/\left(f/f_0\right)^{1/3} \quad (18)$$

where, $\dot{m}$ and $U_i$ are respectively the inlet mass flow rate, and velocity, $Q$ is total heat rate, $A_{ht}$ is the area of heated wall, and $k_m$ and $c_{p,m}$ are thermal conductivity and specific heat capacity of fluid at the arithmetic mean temperature of the inlet and outlet ($\overline{T_i} + \overline{T_o}/2$). In the definition of $\eta$ the subscript "$0$" belongs to the circular tube properties.

According to the obtained velocity and temperature distribution of the flow field, the volumetric entropy generation rate ($\dot{S}_g'''$) related to irreversibility caused by flow friction ($\dot{S}_{g,\Delta p}'''$) and heat transfer ($\dot{S}_{g,\Delta T}'''$) could be expressed as follows [26]:

$$\dot{S}_g''' = \dot{S}_{g,\Delta p}''' + \dot{S}_{g,\Delta T}''' \quad (19)$$

$$\dot{S}_{g,\Delta p}''' = \frac{\mu}{T}\left(\frac{\partial u_i}{\partial x_j} + \frac{\partial u_j}{\partial x_i}\right)\frac{\partial u_i}{\partial x_j} \quad (20)$$

$$\dot{S}_{g,\Delta T}''' = \frac{k}{T^2}\left[\left(\frac{\partial T}{\partial x}\right)^2 + \left(\frac{\partial T}{\partial y}\right)^2 + \left(\frac{\partial T}{\partial z}\right)^2\right] \quad (21)$$

Bejan number (*Be*), as a non-dimensional parameter, is considered to evaluate the influence of the heat transfer entropy generation to the total entropy generation. Moreover, the dimensionless total entropy generation number ($S_N$) is expressed as follows.

$$Be = \frac{\dot{S}_{g,\Delta T}'''}{\dot{S}_g'''} \quad (22)$$

$$S_N = \frac{T_i \iiint_V \dot{S}_g'''}{Q} \quad (23)$$

where *V* is total volume of the heated zone.

## 3. Results and discussions

### 3.1. Grid and inlet-length study

Independence of the computational results from the spatial discretization is tested via successive refinement of non-uniform patterned hexahedral volume elements and comparison of the mean Nusselt number and the apparent friction factor at each refinement step.

Five different grid sizes consisted of 95,400 (Very coarse), 200,800 (Coarse), 337,000 (Intermediate), 566,800 (Fine) and 772,850 (Very fine) elements are considered. The results of grid study for a twisted tube with an *AR* of 2.0 at *Re*=800 are presented in Table 2. Ultimately, by a tradeoff between time, cost and required accuracy, the "fine" grid with 566,800 computational cells was chosen.

To approach a fully developed flow condition at the inlet of the twisted zone, four different lengths of the inlet zones (5, 10, 15 and 20 times of the tube hydraulic diameter) are investigated. The effects of the inlet zone length on the mean Nusselt number and the apparent friction factor for a twisted tube with different aspect ratios at *Re*=800 are illustrated in Figure 2, and the inlet zone length of $15D_h$ is selected for the rest of the calculations.

### 3.2. Code validation

The accuracy and reliability of the thus-utilized solver is verified through the numerical simulation of the laminar flow and heat transfer in a smooth circular tube. Detailed information about test cases can be found in [27]. The obtained results for Nusselt number are compared with the predictions of Sieder-Tate correlation [24, 27] (Eq. 24) and experimental results of Yang et al. [7] in Figure 3a.

$$Nu = 1.86 Pr_f^{1/3} Re^{1/3} \left(\frac{D_h}{L}\right)^{1/3} \left(\frac{\mu_f}{\mu_w}\right)^{0.14} \quad (24)$$

where subscripts of "*f*" and "*w*" represent the fluid and water, respectively.

Figure 3b shows a plot where the numerical results of the friction factor are compared to the Darcy friction formulae [23] (Eq. 25).

$$f = 64/Re \quad (25)$$

Our numerical results show a satisfactory agreement with the Sieder-Tate correlation, Darcy friction factor and experimental results of Yang et al. [7] with lower than 1% average deviation.

### 3.3. Fluid flow pattern

Simulations are performed for various *Re* ranging from 500 to 1100. Contours of non-dimensional velocity ($U/U_i$) and secondary flow vectors are presented in Figure 4 at $z/D_h$=30 of the tube for *Re* equal to 500, 800 and 1100. Clearly, secondary flow does not exist in the circular tube, whereas in twisted oval tubes, they are emerged due to a body force provided by the twisted wall. It is also observed that a bigger tube *AR* leads to a stronger secondary flow and consequently more intensive turbulence as well as effective mixing similar to what already reported in the experimental results of Yang et al. [7]. It is, in addition, found that the secondary flow of the case with an *AR* of 1.5 is quite different from cases with *AR*s of 2.0 and 2.5. In the former, the secondary flow is mostly transitive and perpendicular to the major axis of the cross section, while it is in the form of spiral flow along the tube axis of the latter cases. It is mainly caused by variations in geometrical characteristics of the tubes in compared cases. Despite the low *AR* case with no spiral flow, in cases with higher *AR*s the axially-helical curvature imposes a stronger centrifugal force developing a spiral stream with a stronger secondary flow. It is also observed that induced vortex in the tube cross section makes the fluid to churn, which results in a tendency of the near-the-wall fluid to drift towards the center and augment the secondary flow. For the elliptical tubes, a secondary flow with higher velocity is generated at the proximity of the tipping point. We have also observed that the cross-section circulation is up to 5 fold higher in twisted oval tubes compared to circular tubes and is directly related to the tube *AR* increase.

Figure 5 shows the impact of *Re* on the product of friction factor and *Re* (*fRe*) for the circular tube compared to three different configurations of the twisted oval tubes. Higher pressure drop is noticed in twisted tubes as a result of increase in the tube wall surface area as well as the intensified secondary flow. For the entire range of *Re* studied here, more pressure drop for tubes with twisted wall is observed compared to the circular tube. It is observed that *fRe* increases approximately about 7~10% for twisted oval tubes with different *AR*s, compared to the smooth circular tube at *Re* equal to 500. Furthermore, compared to the smooth circular tube at *Re* of 1100, the *fRe* product increases slightly more (about 17, 18 and 26% for *AR*s of 1.5, 2.0 and 2.5, respectively). The different slope for the case with an *AR* of 1.5 is suspected to stem from its already-presented distinct secondary flow pattern (See Figure 4). It is ultimately observed that at low *Re* by decreasing the cross-sectional *AR* of the tube, viscous effects dominate the centrifugal force; while at higher *Re*, increasing the cross sectional *AR* of the tube would increase the core swirl due to higher centrifugal force. Therefore, it can be explained that by observing higher fluid temperature in tubes with higher *AR*s, lower fluid viscosity is detected at a same Reynolds number. Additionally, the bulk flow temperature is decreased by increasing the Reynolds number which results in higher fluid viscosity. At low Reynolds numbers the concurrent effects of higher viscosity and presence of secondary flow introduce higher pressure drop for tube with *AR*=1.5; while, by increasing the Reynolds number and overcoming the effects of centrifugal forces to viscous forces higher pressure penalty is seen due to the stronger secondary flow in tubes with higher *AR*s.

### 3.4. Heat transfer

Heat transfer performance of the twisted oval tubes is investigated in this section by considering the results of Nusselt number (*Nu*).

The secondary flow, induced by the twisted-wall generated vortices, changes the thermal boundary-layer thickness, intensifies the turbulence within the tube and also enhances the mixing of the hot and cold fluid adjacent to the tube wall and center. As already mentioned (Figure 4), the secondary flow is more potent at higher *AR*s and, on account of greater mixing, results in a more intensified heat transfer in twisted oval tubes. Variations of $Nu_m$ with *Re* are shown in Figure 6, wherein a considerable increase of $Nu_m$ is observed as a result of *AR* rise in twisted oval tubes compared to circular tubes. Although similarly more pronounced heat transfer performance is observed as the *Re* and *AR* increase. The rise of $Nu_m$, and consequently enhancement of heat transfer performance, by increasing the *Re* is suspected to be due to the convection promotion. In addition, the mixing enhancement caused by twisting the tube walls maintains the high temperature gradient near the tube wall and subsequently results in improved heat transfer performance.

Figure 7 presents the variations of the *Nu* along the main flow direction for all configurations at *Re* of 800. Similarly, higher *Nu* is observed for twisted oval tubes compared to the circular one, where the highest value is seen at the topmost value of the cross-section *AR* because of more potent secondary flow and higher temperature gradients. It is also observed that *Nu* decreases along the main flow direction independent of the cross-section *AR*, whereas it is gradual and linear for the circular tube and in fluctuating form for twisted oval tubes. This phenomenon may occur due to the influence of centrifugal force, secondary flow intensity change along the main flow direction inside the tube as well as the interaction of Dean vortices and tube wall, which disturbs the fluid path in twisted oval tubes.

As expected, promoting of the secondary flow intensity by twisting the oval tube wall significantly influences the temperature field passing through the twisted oval tube. Nevertheless, the overall performance of this technique for heat transfer enhancement depends on both heat transfer and working fluid characteristics.

### 3.5. Overall performance

In order to investigate the thermo-hydraulic performances of tubes with different configurations, a performance evaluation criterion, $\eta$ proposed by Webb [25], is chosen as a measure of the heat transfer enhancement with the expense of the pumping power increase. Thus, the higher the value of $\eta$, the better overall performance of the tube. Figure 8 shows variations of $\eta$ versus *Re* for the circular tube and three different configurations of the twisted oval tube. It is observed that performance factor is higher than that of the circular tube over the entire range of the *Re*. It is also observed that overall performance of twisted tubes is significantly enhanced as the *Re* increases.

### 3.6. Field synergy analysis

In 1981, a novel concept called "field synergy principle" was firstly proposed by Guo et al. [28] for enhancement of parabolic flow convective heat transfer. This concept is principally based on the idea that the reduction of the intersection angle between the velocity vector and the temperature gradient, called "synergy angle", can significantly promote convective heat transfer. This concept was extended for elliptic flows by Tao et al. [29] using theoretical analysis and verified numerically in [30]. Guo et al. [31] provided a comprehensive review on recent investigations of field synergy principle and Chen et al. [32] proposed a novel type of heat transfer tube based on field synergy principle.

The synergy angle is also studied here to analyze the heat transfer and fluid flow characteristics of twisted oval tube. The local intersection angle between the velocity vector and the temperature gradient ($\xi$) is defined by Eq. 26 at grid nodes.

$$\xi = \arccos\left(\frac{\mathbf{U}.\nabla T}{|\mathbf{U}||\nabla T|}\right) \quad (26)$$

The value of $\xi$ is an indication of heat transfer intensity, where the worst synergy condition is when the flow streamlines are perpendicular to the temperature gradient. For this situation, the convective heat transfer is not affected by the fluid velocity, or in another word it is independent from Reynolds number. When the $\xi$ approaches 0º, the flow configuration leads to heating of fluid and the flow configuration is cooling the fluid as the $\xi$ approaches to 180º.

Figure 9 shows the variations of volume-averaged synergy angle ($\xi_m$) with *Re*. It is shown that the value of $\xi_m$ for the TOTs decreases with the increase of *Re*; whilst it is almost independent of the *Re* for circular tube. For the circular tube there is no secondary flow, so the trend of Volume-averaged synergy angle is different from that of twisted oval tubes. At higher Reynolds numbers the impact of fluid axial conduction become weaker and therefore the synergy angle is increased. It is indeed noticeable that the value of $\xi_m$ for the twisted oval tubes is smaller than that of the circular tube, which is consistent with the fact that the $Nu_m$ is higher for them compared to circular tubes (see Figure 6). It can be explained that the vortices induced by the twisted walls improve the synergy between velocity vectors and temperature gradient, increasing the wall velocity gradient and decreasing the thermal boundary layer thickness and results in heat transfer enhancement. Compared to the circular tube, the twisted oval tubes with higher cross-section *AR*s have smaller values of $\xi_m$ as well as higher *Nu* in the range of the *Re* studied in the present paper, thus a better heat transfer performance is achieved as expected.

### 3.7. Entropy generation analysis

Any heat transfer enhancement techniques will cause higher pressure penalty which may result higher entropy generation. Non-dimensional entropy generation ($S_N$) as function of *Re*, for different configurations is shown in Figure 10. It is clear that $S_N$ tends to decrease with an increase in *Re* for TOTs, while, $S_N$ slightly increases for the circular tube. It can be explained that for the circular tube the bulk flow temperature is lower at higher Reynolds numbers which results in higher fluid viscosity and also higher velocity gradients. These effects introduce higher irreversibility contributed by fluid friction and lessen the irreversibility caused by heat transfer. It is observed that the non-dimensional total entropy generation declines as the tube cross sectional aspect ratio increases at the same Reynolds number. To evaluate the relative influence of flow friction and heat transfer irreversibility on total entropy generation, the variations of the Bejan number (*Be*) with Reynolds number are illustrated in Figure 11 for different configurations. It is detected that at the higher *Re*, Bejan number tends to decrease which shows that the influence of the entropy generation contributed by flow friction in the total entropy generation increases and the effect of heat transfer on total entropy generation decreases. It is worth mentioning that in Figure 11 all the Bejan numbers are close to one, which means that the effect of flow friction on the total entropy generation is insignificant. Among all the mentioned configurations, tube with an *AR* of 1.5 has the lowest Bejan number for the entire range of the *Re* which shows that the entropy generation contributed by the flow friction is the highest compared to that of the other configurations. It is found that the influence of heat transfer on total entropy generation increases as the *AR* increases at the same Reynolds number.

As discussed by Guo et al. [33], for most liquid flows the irreversibility caused by the flow friction is much smaller than that because of heat transfer and hence, no exterma may exist for non-dimensional total entropy generation function for the cases investigated in this study.

## 4. Conclusions

Three dimensional numerical study of the flow pattern and heat transfer performance of twisted oval tubes were conducted for *Re* in the range of 500 to 1100. The effects of the geometric characteristics and *Re* change on the performance of the twisted oval tubes were investigated and the overall performance of the twisted oval tubes was compared to the circular tube. The field synergy principle and entropy generation analysis were also utilized to explore heat transfer enhancement mechanism, and the following major findings and conclusions are obtained.

The higher heat transfer rate in twisted oval tubes is the direct result of enhanced fluid mixing caused by the induced vortices inside the tube. The secondary flow is in the form of spiral flow along the tube axis for high *AR*s, while for low *AR*s, it is mostly transitive and perpendicular to the cross sectional major axis. The enhanced *AR*-related secondary flow in twisted tubes brings the hotter fluid from the walls toward the central region and vice versa, and increases the temperature gradients at the walls; thus enhancing the heat transfer inside the tubes. The results indicate that heat transfer enhancement arising from the improved synergy between velocity vectors and temperature gradient.

The surface area increase and also intensifying the secondary flow in twisted oval tubes would lead to higher pressure drop compared to the circular tubes and is directly related to the *AR* of the tube cross section.

Twisted oval tubes have better heat transfer performance compared to circular tubes, wherein increasing the *Re* and the tube cross section *AR* will increase the heat transfer coefficient. Elevated heat transfer performance is also expected in cases where fluids with higher viscosity and thermal conductivity are utilized.

The increment of pumping power dominates the increment of the heat transfer rate with increasing the *Re*. Thus, for space-constrained applications where the pumping power is not an important design parameter, twisted oval tubes can be a suitable option.

From both field synergy principle and entropy generation analysis, it is observed that all the configurations of TOTs considered in this study perform better performance compared to circular tube. Additionally, the twisted tube with cross section *AR* of 2.5 appears the most promising configuration followed by the twisted tube with *AR*s of 2.0 then 1.5.

All in all, the advantages of using twisted oval tubes related to energy saving elevate as *Re* or cross section *AR* increases. Hence, increasing the turbulence intensity by twisting the tube walls could significantly enhance the thermo-hydraulic performances of tubes.

The enhanced heat transfer performance is a direct result of the secondary flow and consequently augmented convective heat transfer, which has not been respected enough so far. Additional heat transfer studies in chorus with fluid dynamics are recommended to offer novel designs of the heat exchangers with twisted tubes.

NOMENCLATURE

| | |
|---|---|
| A | Area, $m^2$ |
| a | Major axis of tube cross section, *m* |
| AR | Aspect ratio = a/b |
| b | Minor axis of tube cross section, *m* |
| Be | Bejan number |
| $c_p$ | Specific heat capacity, *J/kg.K* |
| $D_h$ | Hydraulic diameter, *m* |
| f | Apparent friction factor |
| h | Convection heat transfer coefficient, $W/m^2 K$ |
| i | Inlet zone length, *m* |
| j | Outlet zone length, *m* |
| k | Thermal conductivity, *W/mK* |
| L | Heated zone length, *m* |
| Nu | Nusselt number |
| p | Pressure, *Pa* |
| Pr | Prandtl number |
| Q | Heat transfer rate, *W* |
| Re | Reynolds number |
| S | Twisted zone length, *m* |
| $S_N$ | Dimensionless total entropy generation |
| T | Temperature, *K* |
| TOT | Twisted oval tube |
| TOTHE | Twisted oval tube heat exchanger |
| U | Fluid velocity vector |
| u, v, w | Velocity vector components |
| V | Total volume of the heated zone, $m^3$ |
| x, y, z | Cartezian coordinates |
| $\dot{m}$ | Mass flow rate, $kg/m^3$ |

Greek Symbols

| | |
|---|---|
| η | Performance factor |
| μ | Dynamic viscosity, *kg/m.sec* |
| ξ | Synergy angle, ° |
| ρ | Fluid density, $kg/m^3$ |

Subscripts

| | |
|---|---|
| 0 | Circular tube |
| f | Fluid |
| ht | Heated |
| i | Inlet |
| m | Mean |
| o | Outlet |
| w | Water |
| wall | Wall |

List of Figures

Figure 1- Physical model and relevant geometrical parameters of twisted tubes.

Figure 2- The effects of the inlet zone length on the mean Nusselt number and the apparent friction factor for TOTs with different aspect ratios at $Re=800$.

Figure 3- Comparison of (a) the Nusselt number with available experimental data and theoretical results (b) the friction factor with Darcy friction factor for different Reynolds numbers.

Figure 4- The contours of non-dimensional velocity and the secondary flow vectors at different cross sections of tube ($Re=800$).

Figure 5- Effects of different configurations and Reynolds number on product of friction factor and $Re$ ($fRe$).

Figure 6- The relationships of mean Nusselt number ($Nu_m$) with Reynolds number for different configurations.

Figure 7- Distribution of Nusselt number in the heated zone for different configurations ($Re=800$).

Figure 8- Effect of different configurations on overall performance for different Reynolds numbers.

Figure 9- Volume-averaged synergy angle versus Reynolds number for different configurations.

Figure 10- The non-dimensional total entropy generation versus Reynolds number for different configurations.

Figure 11- Variations of Bejan number ($Be$), as a function of Reynolds number for different configurations.

List of tables

Table 1- The effective temperature-dependent thermo-physical properties of water.

Table 2- The results of grid independency test for a twisted tube with an *AR* of 2.0 at *Re*=800.

Highlights

- Laminar fluid flow and heat transfer in twisted oval tubes is numerically investigated.
- Liquid flow, heat transfer and entropy generation in twisted oval tubes are scrutinized.
- The results of simulations are analyzed in view-point of the field synergy principle.
- The overall efficiency of twisted oval tubes is studied

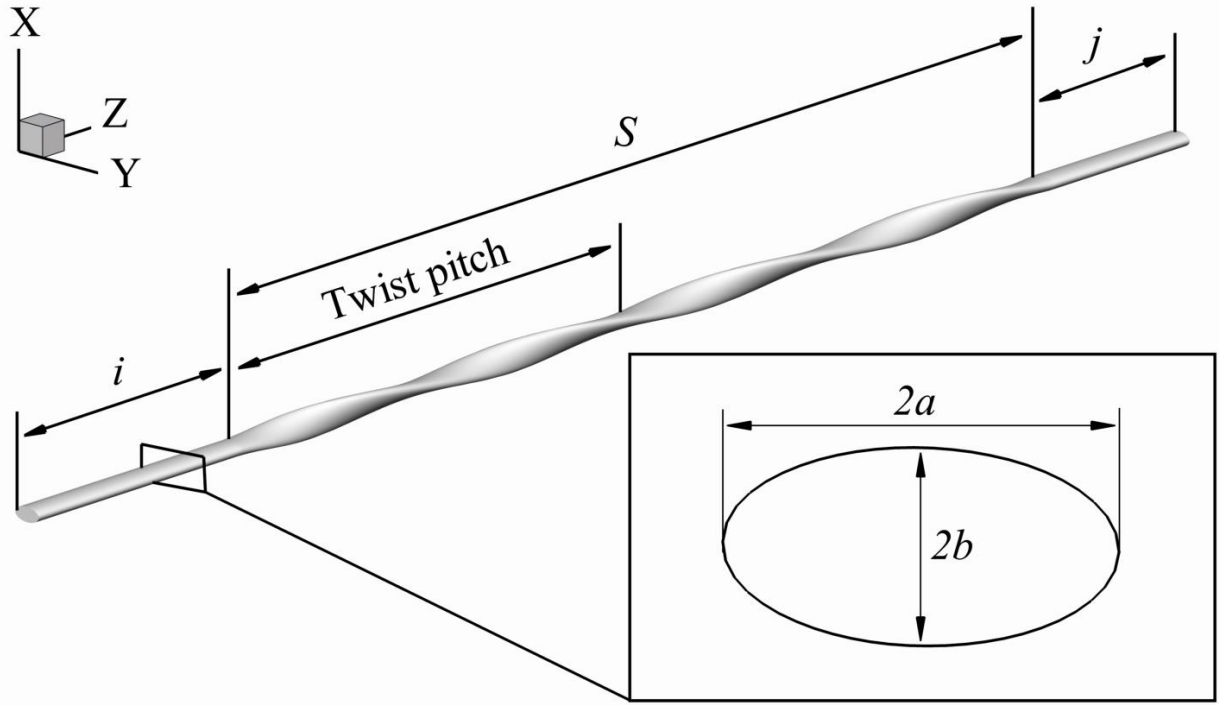

Figure 1- Physical model and relevant geometrical parameters of twisted tubes.

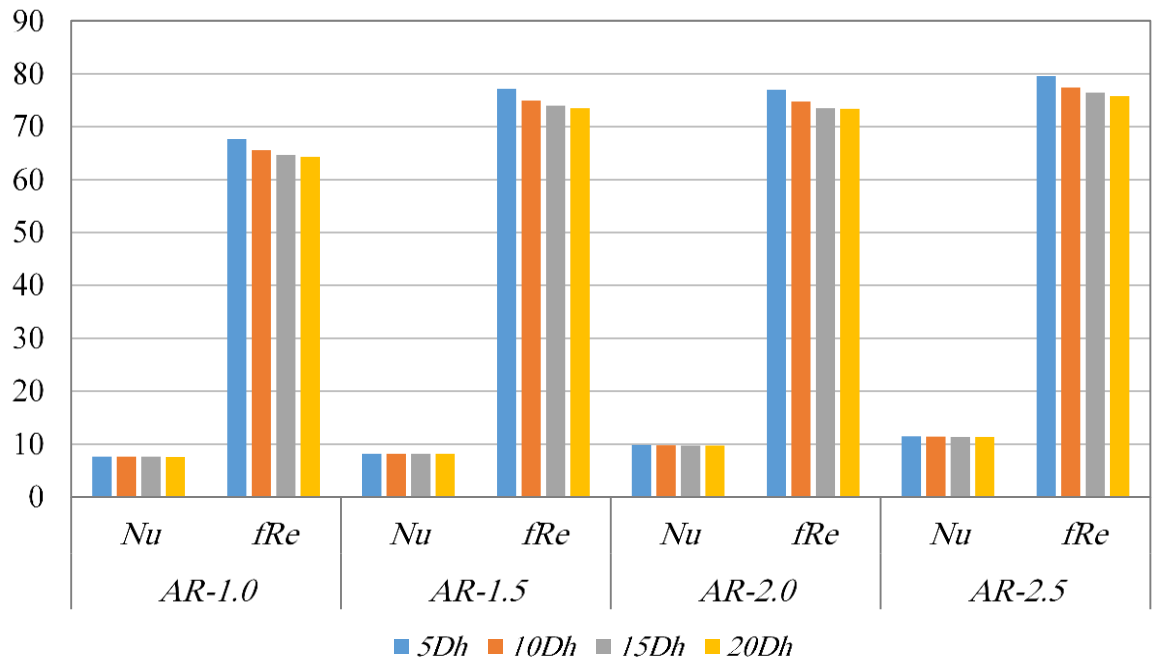

Figure 2- The effects of the inlet zone length on the mean Nusselt number and the apparent friction factor for TOTs with different aspect ratios at *Re*=800.

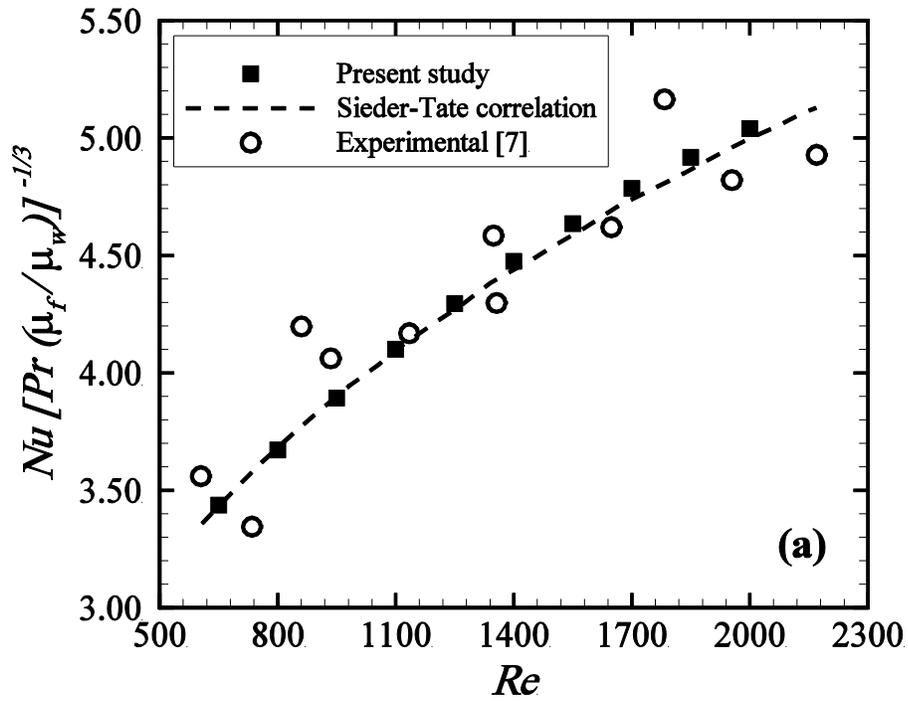

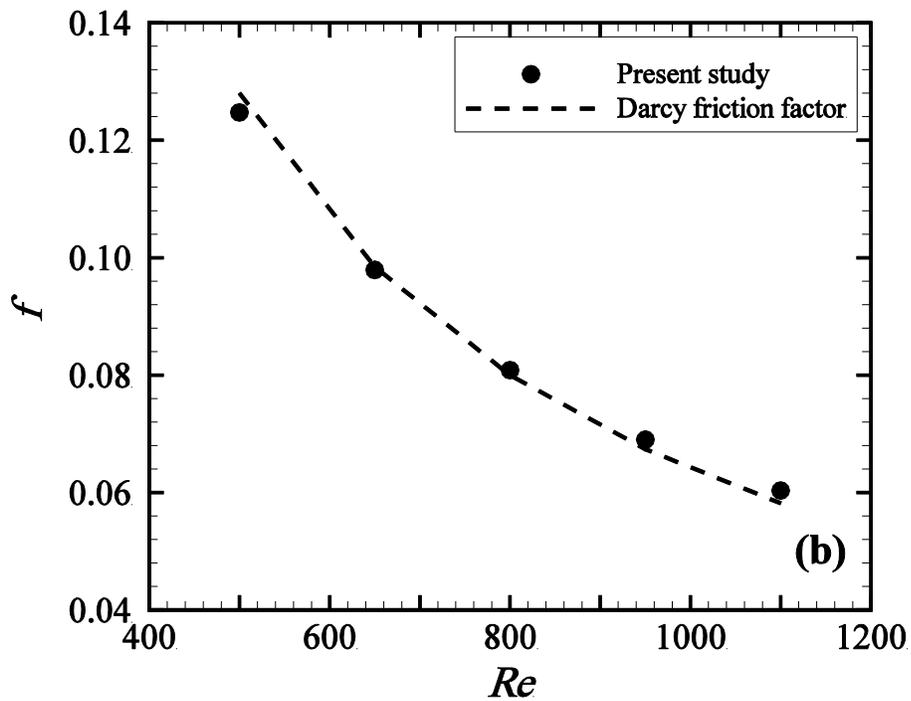

Figure 3- Comparison of (a) the Nusselt number with available experimental data and theoretical results (b) the friction factor with Darcy friction factor for different Reynolds numbers.

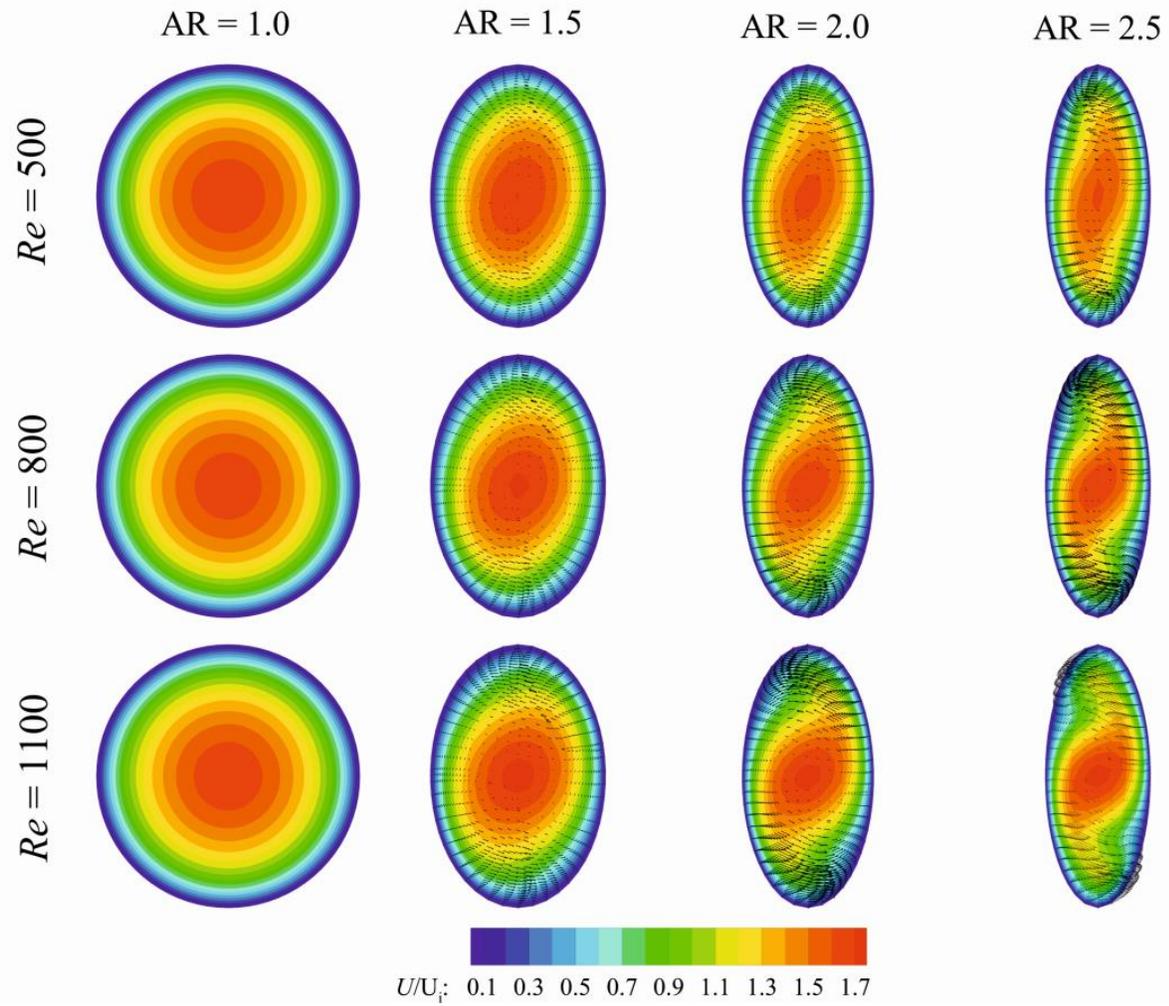

Figure 4- The contours of non-dimensional velocity and the secondary flow vectors at different cross sections of tube (*Re*=800).

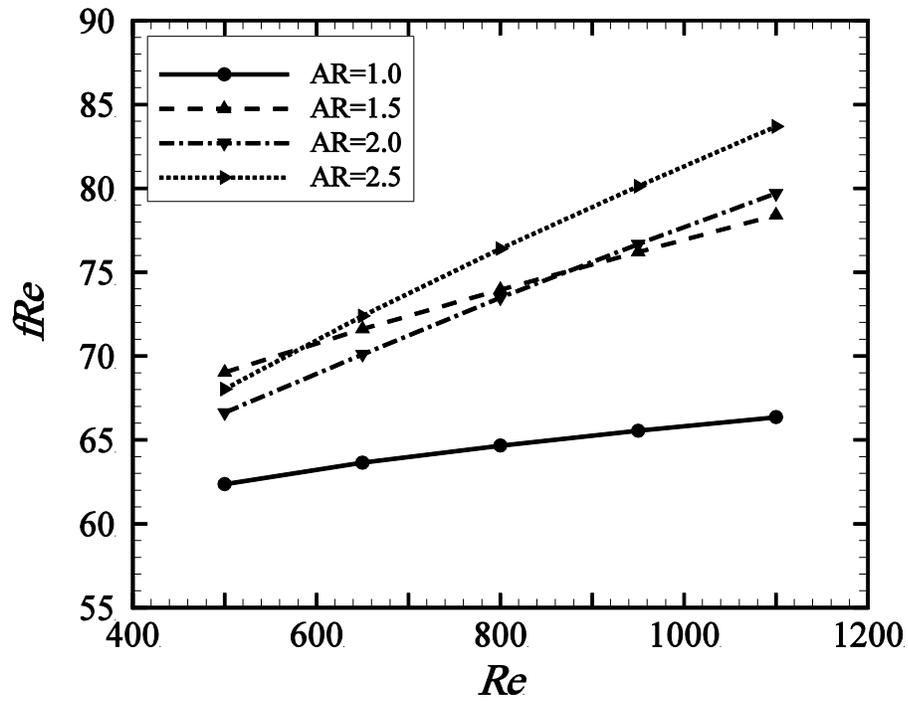

Figure 5- Effects of different configurations and Reynolds number on product of friction factor and *Re* (*fRe*).

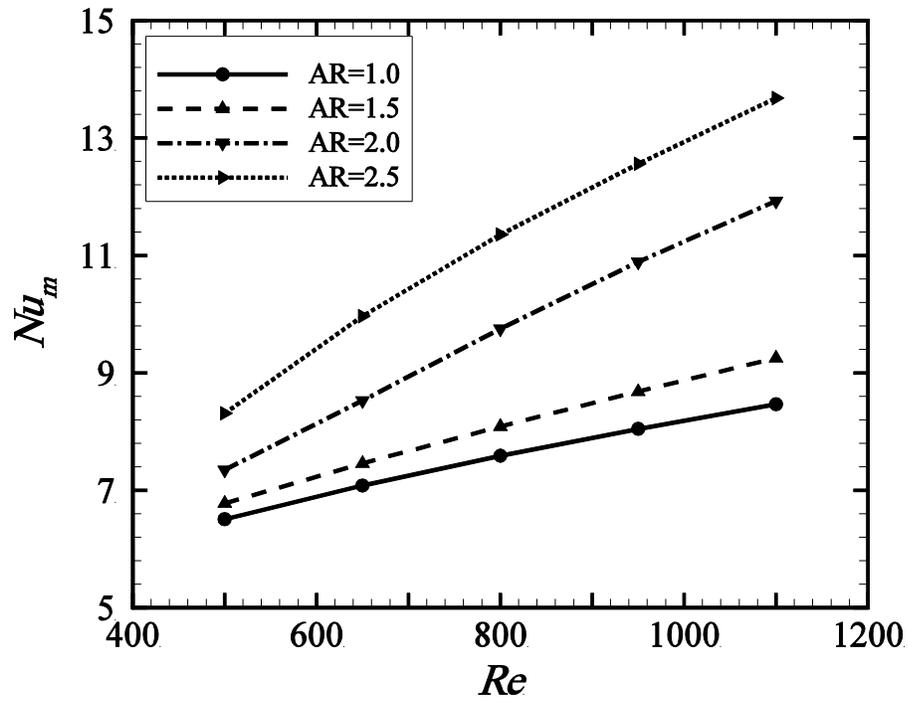

Figure 6- The relationships of mean Nusselt number ($Nu_m$) with Reynolds number for different configurations.

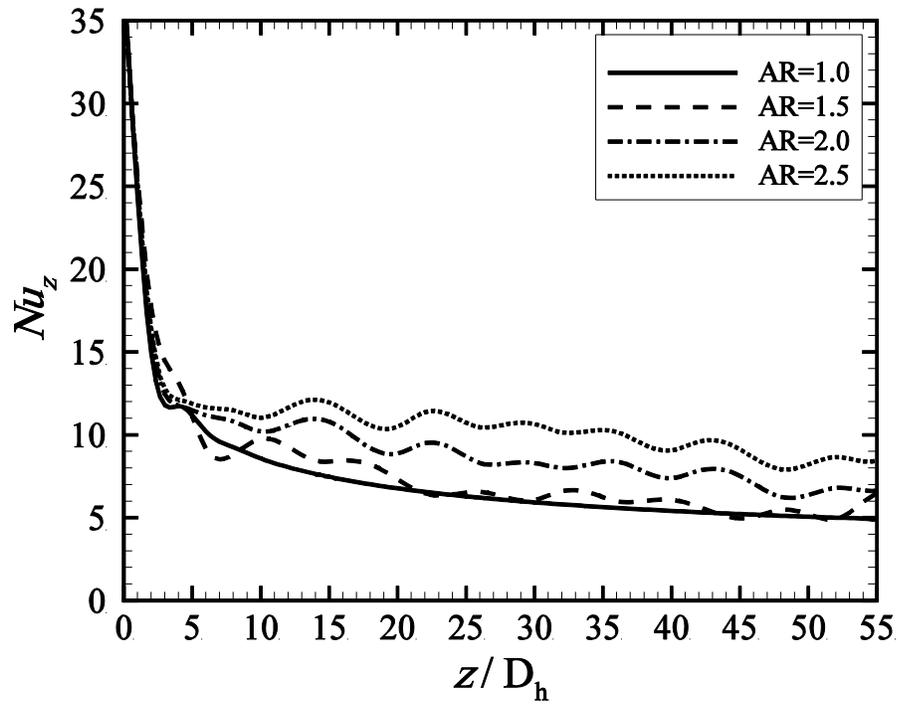

Figure 7- Distribution of Nusselt number in the heated zone for different configurations (*Re*=800).

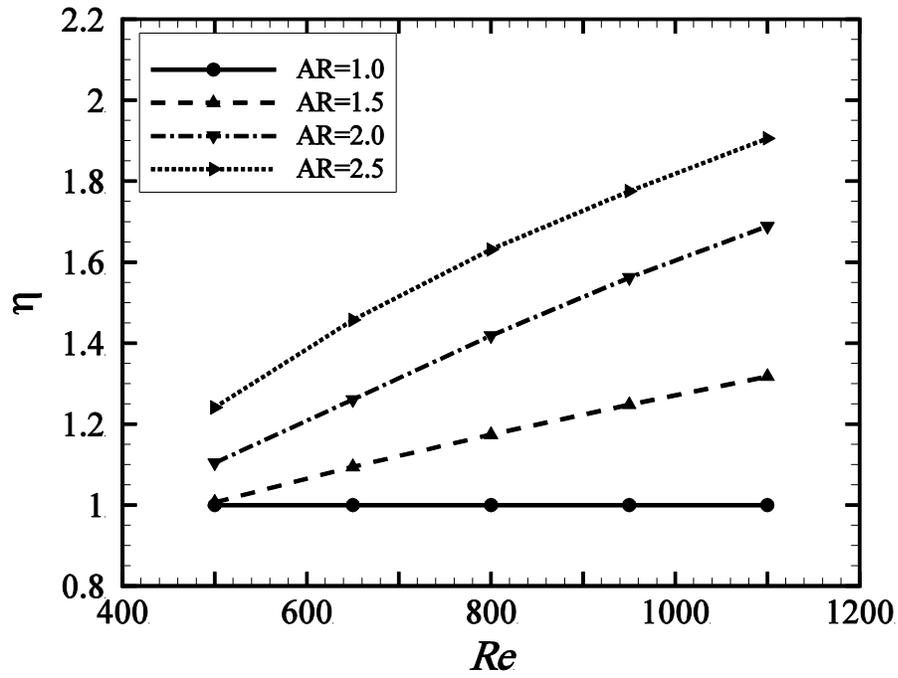

Figure 8- Effect of different configurations on overall performance for different Reynolds numbers.

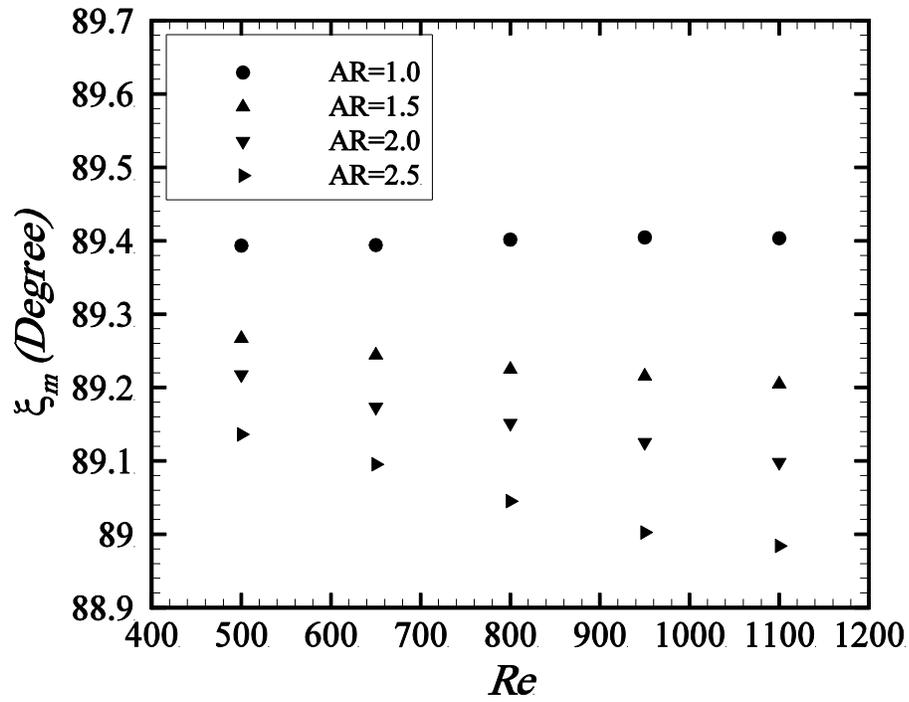

Figure 9- Volume-averaged synergy angle versus Reynolds number for different configurations.

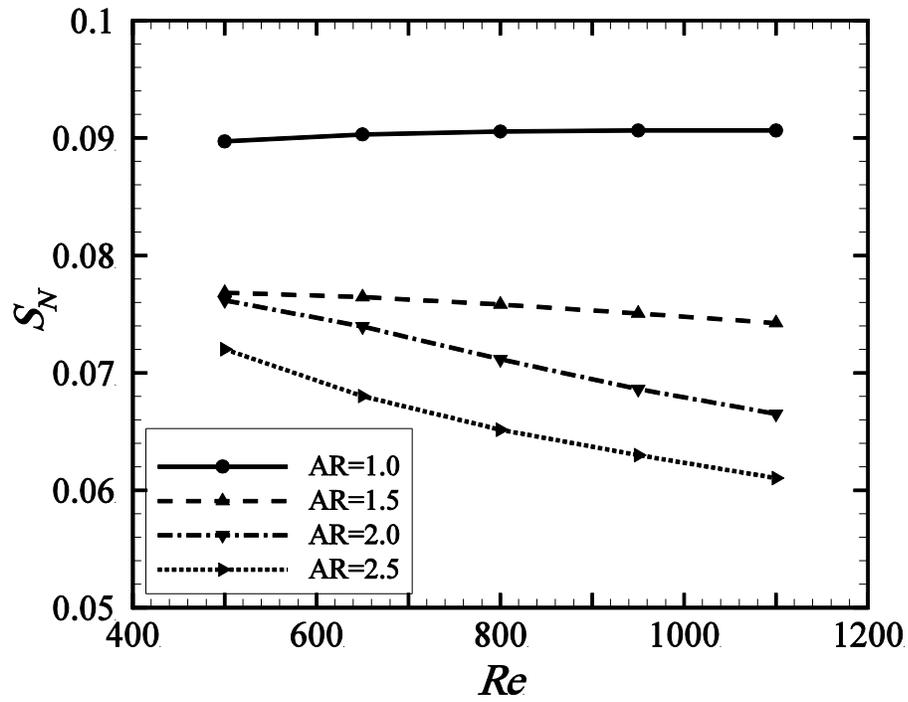

Figure 10- The non-dimensional total entropy generation versus Reynolds number for different configurations.

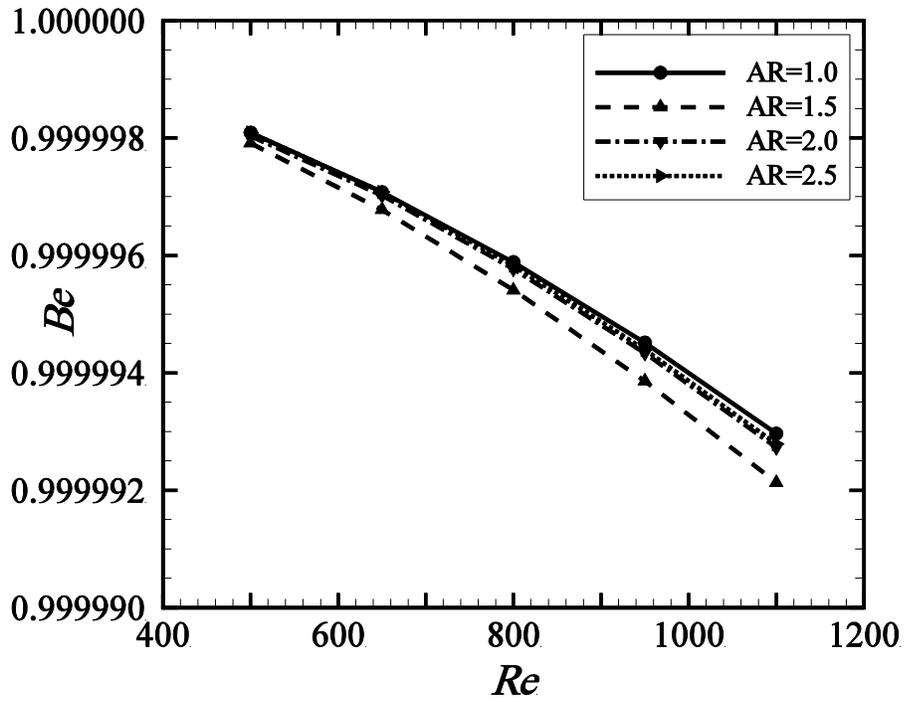

Figure 11- Variations of Bejan number (*Be*), as a function of Reynolds number for different configurations.

Table 1- The effective temperature-dependent thermo-physical properties of water.

| Properties [24, 34] | Temperature dependent function |
| --- | --- |
| $\rho_{Water}$ | $753.2 + 1.88T - 3.570 \times 10^{-3} T^2$ |
| $k_{Water}$ | $-0.5981 + 6.53 \times 10^{-3} T - 8.354 \times 10^{-6} T^2$ |
| $\mu_{Water}$ | $2.591 \times 10^{-5 + (238.3/(T - 143.2))}$ |
| $c_{p,Water}$ | 4200 |

Table 2- The results of grid independency test for a twisted tube with an *AR* of 2.0 at *Re*=800.

| Number of Cells | $Nu_m$ | % Diff $Nu_m$ | $f$Re | % Diff $f$Re |
| --- | --- | --- | --- | --- |
| 95400 | 10.115 | - | 74.908 | - |
| 200800 | 9.850 | -2.618 | 74.491 | -0.557 |
| 337000 | 9.767 | -0.842 | 73.643 | -1.138 |
| 566800 | 9.756 | -0.109 | 73.488 | -0.211 |
| 772850 | 9.753 | -0.034 | 73.397 | -0.125 |